\documentclass{aa}

\usepackage{txfonts}
\usepackage{graphicx}

\begin{document}

\title{Analytical evaluation of the X-ray scattering contribution\\
 to imaging degradation in grazing-incidence X-ray telescopes}

\author{D. Spiga}

\institute{INAF/Osservatorio Astronomico di Brera,
  Via E. Bianchi 46, I-23807, Merate (LC) - Italy}

\offprints{daniele.spiga@brera.inaf.it}

\date{Received 02 February 2007 / Accepted 27 March 2007}

\abstract 
{}
{The focusing performance of X-ray optics (conveniently expressed in terms of HEW, Half Energy Width) strongly depend on both mirrors deformations and  photon scattering caused by the microroughness of reflecting surfaces. In particular, the contribution of X-ray Scattering (XRS) to the HEW of the optic is usually an increasing function $H(E)$ of the photon energy $E$. Therefore, in future hard X-ray imaging telescopes of the future (SIMBOL-X, NeXT, Constellation-X, XEUS), the X-ray scattering could be the dominant problem since they will operate also in the hard X-ray band (i.e. beyond 10 keV). In order to ensure the imaging quality at all energies, clear requirements have to be established in terms of reflecting surfaces microroughness.}
{Several methods were proposed in the past years to estimate the scattering contribution to the HEW, dealing with the surface microroughness expressed in terms of its Power Spectral Density (PSD), on the basis of the well-established theory of X-ray scattering from rough surfaces. We faced that problem on the basis on the same theory, but we tried a new approach: the direct, analytical translation of a given surface roughness PSD into a $H(E)$ trend, and -- vice versa -- the direct translation of a $H(E)$ requirement into a surface PSD. This PSD represents the maximum tolerable microroughness level in order to meet the $H(E)$ requirement in the energy band of a given X-ray telescope.}
{We have thereby found a new, analytical and widely applicable formalism to compute the XRS contribution to the HEW from the surface PSD, provided that the PSD had been measured in a wide range of spatial frequencies. The inverse problem was also solved, allowing the immediate evaluation of the mirror surface PSD from a measured function $H(E)$. The same formalism allows establishing the maximum allowed PSD of the mirror in order to fulfill a given $H(E)$ requirement. Practical equations are firstly developed for the case of a single-reflection optic with a single-layer reflective coating, and then extended to an optical system with $N$ identical reflections. The results are approximately valid also for multilayer-coated mirrors to be adopted in hard X-rays. These results will be extremely useful in order to establish the surface finishing requirements for the optics of future X-ray telescopes.} 
{}

\titlerunning{Analytical evaluation of X-ray scattering contribution to imaging degradation $\ldots$}
\authorrunning{D. Spiga}
\maketitle

\section{Introduction}

The adoption of grazing-incidence optics in X-ray telescopes in the late 70s allowed a great leap forward in X-ray astronomy because they endowed the X-ray instrumentation with imaging capabilities in the soft X-ray band ($E <$ 10~keV). The excellent performances of the soft X-ray telescopes ROSAT (Aschenbach~\cite{AschenbachROSAT}), Chandra (Weisskopf~\cite{Weisskopf}) and Newton-XMM (Gondoin~et~al.~\cite{Gondoin}) are well known.

To date, the utilized technique to focus soft X-rays consists in systems of double-reflection mirrors with a single layer coating (Au, Ir) in total external reflection at shallow grazing incidence angles. In this case, the incidence angle $\theta_{\mathrm i}$ (as measured from the mirror surface) cannot exceed the critical angle for total reflection, otherwise the mirror reflectivity would be very low. The critical angle is inversely proportional to $E$, the energy of the photons to be focused. Using Au coatings, for instance, the incidence angle cannot exceed $\sim$~0.4~deg for photon energies $E \approx$~10~keV.

 An extension of this technique to the hard X-ray energy band ($E>$~10~keV) can be pursued by combining long focal lengths ($>$~10~m), very small incidence angles (0.1~$\div$~0.25~deg), and wideband multilayer coatings to enhance the reflectance of the mirrors at high energies (Joensen~et~al.~\cite{Joensen}; Tawara~et~al.~\cite{Tawara}). A very long focal length is hardly managed using a single spacecraft, therefore the optics and the focal plane instruments should be carried by two separate spacecrafts in formation-flight configuration. This is the baseline for the future X-ray telescopes SIMBOL-X (Pareschi~\&~Ferrando~\cite{Ferrando}) and XEUS (Parmar~et~al.~\cite{Parmar}). Other hard X-ray imaging telescopes of the future are NeXT (Ogasaka~et~al.~\cite{Ogasaka}) and Constellation-X (Petre~et~al.~\cite{Petre}).

The focusing and reflection efficiency of X-ray optics can be tested and calibrated on ground by means of full-illumination X-ray facilities like PANTER (Br\"auninger~et~al.~\cite{Brauninger}; Freyberg~et~al.~\cite{Freyberg}), successfully utilized in the last years to calibrate the optics of a number of soft X-ray telescopes. The PANTER X-ray facility now allows testing in soft (0.2~$\div$~10~keV) and hard (15~$\div$~50~keV) X-rays multilayer-coated optics prototypes for future X-ray telescopes (Pareschi~et~al.~\cite{Pareschi}; Romaine~et~al.~\cite{Romaine}). The source distance finiteness causes some departures of the optic performances, with respect to the case with the source at astronomical distance: effective area loss, different incidence angles on paraboloid and hyperboloid, focal length displacement, a slight focal spot blurring (Van Speybroeck~\&~Chase~\cite{VanSpeyebroeck}). However, there effects can be quantified and subtracted from experimental data. After this treatment, the focusing-concentration performances of the optic can be experimentally characterized as a function of the incident photon energy, in terms of Half-Energy Width (HEW) and Effective Area (EA).

The focusing performance, in particular, is altered by mirror deformations that may arise in the manufacturing, handling, integration, positioning processes. The consequent imaging degradation can be calculated from the measured departures of the mirrors from the nominal profile, by means of a ray-tracing program. As long as the geometrical optics approximation can be applied, the effect is independent of the photon energy. The figure errors contribution to the HEW can be also directly measured using a highly collimated beam of visible/UV light in a precision optical bench. In this case, however, the light diffraction has to be carefully estimated and subtracted.

Another drawback is the X-ray scattering (XRS) caused by the microroughness of reflecting surfaces (Church~et~al.~\cite{Church}; Stearns~et~al.~\cite{Stearns}; Stover~\cite{Stover}; and many others). The XRS spreads a variable fraction of the reflected beam intensity in the surrounding directions: the result is the effective area loss in the specular direction (i.e. in the focus) and a degradation of the imaging quality. The XRS is an increasing function of the photon energy; due to the impact that the XRS can have on astronomical X-ray images quality, the height fluctuations rms of the mirror surface should not exceed few angstr\"{o}ms. Loss of effective area is also caused by interdiffusion of layers in multilayer coatings, which enhances the X-ray transmission and absorption throughout the stack. On the other hand, an uniform interdiffusion does not cause X-ray scattering (Spiller~\cite{Spiller}), hence it does not contribute to the focusing degradation. 

The microroughness of an X-ray mirror can be measured on selected samples using several metrological instruments, each of them sensitive to a definite interval of spatial scales $\hat{l}$: Long Trace Profilometers (10~cm $>\hat{l}>$ 0.5~mm: Tak\'{a}cs~et~al.~\cite{Takacs}), optical interference profilometers (5~mm $>\hat{l}>$ 10~$\mu$m) and Atomic Force Microscopes (100~$\mu$m $>\hat{l}>$ 5~nm) can be suitable instruments to provide a detailed profile characterization of X-ray mirrors surface. It is convenient to present the deviation of surface from the ideality in terms of Power Spectral Density (PSD), because its values do not depend on the measurement technique in use (see \cite{ISO10110}). In addition, the XRS diagram, and consequently the HEW, can be immediately computed from the PSD at any photon energy (Church~et~al.~\cite{Church}). 

In the past years, several approaches were elaborated to relate a mirror PSF (Point Spread Function) to the PSD of its surface. Among a wealth of works, we can cite (De~Korte~et~al.~\cite{DeKorte}) the assumption of a Lorentzian model for the PSD to fit the mirror PSFs at some photon energies, allowing the derivation of two parameters (roughness rms and correlation length) of the model PSD. Christensen~et~al. (\cite{Christensen}) perform a fit of experimental high-resolution XRS data dealing with the surface correlation function. Harvey~et~al. (\cite{Harvey88}) relate the PSF of Wolter-I optics to the parameters of an exponential self-correlation function along with a transfer function-based approach. Willingale (\cite{Willingale}) derived the surface PSD of a mirror from the wings of a few PSFs, measured at PANTER at some soft X-ray photon energies. O'Dell~et~al. (\cite{ODell}) interpret the PSF of a focusing mirror on the basis of surface roughness and particulate contamination. Zhao~\&~Van~Speybroeck (\cite{Zhao}) construct from the PSD of a focusing mirror a model surface and compute the X-ray scattering PSF from the Fraunhofer diffraction theory.

 In the present work that problem is faced in a new and different way, looking for a general and simple link between measured roughness and mirror HEW. More precisely, we considered the following question: \it for an X-ray grazing-incidence optic, what is the maximum acceptable PSD of the surface that fulfills the angular resolution (HEW) requirements of the telescope, in all the energy band of sensitivity\rm?

In this work we shall give a definite answer to this question. In the sect.~\ref{sect:factors} we shall summarize the causes of imaging degradation. In the sect.~\ref{sect:fromPtoH} we show how to evaluate $H(E)$, the XRS contribution to the HEW of a focusing mirror at the photon energy $E$, from {\it any} surface microroughness PSD, measured over a {\it very} wide range of spatial frequencies. We shall see in the sect.~\ref{sect:fractal} that for the special class of {\it fractal surfaces} we can even relate the power-law indexes of PSD and HEW, and in the sect.~\ref{sect:numerical} we see how to treat the other cases. Then we prove in the sect.~\ref{sect:fromHtoPSD} that the formalism can be reversed, providing thereby an independent evaluation of the surface PSD from an analytical calculation over $H(E)$, and in the sect.~\ref{sect:multiple} we extend the results to focusing mirrors with more than one reflection. Finally, an example of computation is provided in the sect.~\ref{sect:example}. 

\section{Contributions to the imaging degradation}\label{sect:factors}

\begin{figure}
	\resizebox{\hsize}{!}{\includegraphics{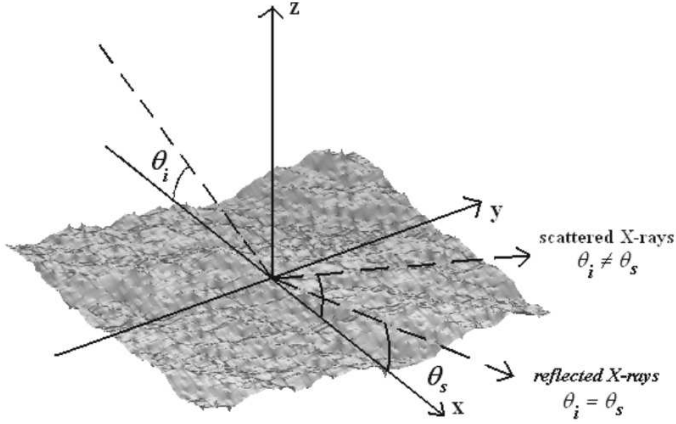}}
	\caption{The geometry of X-ray scattering: the strictly speaking "reflected" rays (i.e. in the focus direction) are characterized by the equality $\theta_s = \theta_i$, the others are scattered apart. The rough surface is a simulated one, assuming a PSD with power-law index $n$~=~2.4 (see sect.~\ref{sect:fractal}).}
\label{fig:XRS}
\end{figure}

We shall henceforth indicate with $\lambda$ the wavelength of photons impinging on the mirror, and we shall consider the HEW as a function of $\lambda$ instead of the photon energy E. For isotropical reflecting surfaces in grazing incidence, the X-ray scattering distribution lies essentially in the incidence plane, so we denote the incidence angle on the mirror as $\theta_{\mathrm i}$ and the scattering angle as $\theta_{\mathrm s}$, both measured from the surface plane (a schematic of the scattering geometry is drawn in fig.~\ref{fig:XRS}). If we do not consider the optic roundness errors, the longitudinal deviations from the nominal profile of a focusing mirror can be classified on the basis of their typical length $\hat{l}$. According to De~Korte~et al.~(\cite{DeKorte}), they are:
\begin{enumerate}
	\item{{\it Power errors}: errors with $\hat{l}$ equal to the mirror length $L$. They consist in a single-concavity deformation of the profile with respect to the nominal one.}
	\item{{\it Regularity errors}: errors in the spatial range from $0.1~L<\hat{l}<0.5~L$. }
	\item{{\it Surface roughness}: surface defects with $\hat{l}<0.1~L$. }
\end{enumerate}

However, other criteria were also formulated to separate figure errors from roughness. Consider a single Fourier component of the surface profile with wavelength $\hat{l}$ and root mean square $\sigma$. That Fourier component is dominated by figure error if it fulfills the condition (Aschenbach~\cite{Aschenbach})
\begin{equation}
	4\pi \sin \theta_{\mathrm i} \sigma > \lambda.
	\label{eq:smooth_surf}
\end{equation}
Otherwise, it is dominated by microroughness. In other words, surface defects within the smooth-surface approximation can be mainly considered as microroughness. To understand the importance of this approximation, we write the optical path difference $\Delta s$ of X-rays reflected by two points of the surface with a horizontal spacing $\hat{l}$ and vertical spacing $\hat{\sigma} = 2\sqrt{2}\sigma $ (for optically-polished surfaces, $\sigma$ is a increasing function of $\hat{l}$, and usually $\sigma \ll 10^{-3} \hat{l}$) as
\begin{equation}
	\Delta s = \hat{l} (\cos \theta_{\mathrm s} - \cos \theta_{\mathrm i})+ \hat{\sigma}(\sin \theta_{\mathrm i} + \sin \theta_{\mathrm s})
	\label{eq:deltas1}
\end{equation}
that, for small incidence angles, becomes
\begin{equation}
	\Delta s = \hat{l}\sin \theta_{\mathrm i} (\theta_{\mathrm s} - \theta_{\mathrm i})+ \hat{\sigma}(\theta_{\mathrm i} + \theta_{\mathrm s}).
	\label{eq:deltas2}
\end{equation}

If that component is responsible for X-ray scattering, it has to be $\Delta s \approx \lambda$, to cause the diffraction from surface features with a $\hat{l}$ spacing and $\hat{\sigma}$ height. Conversely, the "figure errors", which are treated with the methods of the geometrical optics, should be characterized by the inequality $\Delta s \gg \lambda$. Note that this condition becomes similar to the eq. \ref{eq:smooth_surf} in the limit $|\theta_{\mathrm i} - \theta_{\mathrm s}|\rightarrow 0$. The application of this criterion and of the subsequent X-ray scattering theory requires the incident radiation to be spatially coherent over the spatial scale $\hat{l}$, so that the properties of the reflected wavefront are determined only by the coherence properties of the mirror surface. This in turn requires the angular diameter of the source $\phi_{\mathrm S}$ to fulfill the inequality (Hol\'{y}~et~al.~\cite{Holy})
\begin{equation}
	\phi_{\mathrm S} < \frac{\lambda}{\hat{l}\sin\theta_{\mathrm i}}.
\label{eq:ang_diameter}
\end{equation}
This equation sets a maximum to the values of $\hat{l}$ that can be used in the application of the results presented in this work. The limitation can affect X-ray sources at finite distance, like those used for X-ray optics calibrations in full-illumination setup. For very distant astronomical X-ray sources, the condition \ref{eq:ang_diameter} is met even for larger $\hat{l}$, up to $\hat{l} \approx L$. 

It is worth pointing out that, for a given reflecting surface, the separation of figure errors from microroughness is strongly affected by the incidence/scattering angles. In fact, even for large $\hat{l}$, $\Delta s$ can become comparable with $\lambda$, if $\theta_{\mathrm i}$ and $\theta_{\mathrm s}$ are sufficiently small: thus, the spatial wavelength window of interest for X-ray scattering can shift to the large $\hat{l}$ domain (or, equivalently, to the range of low spatial frequencies $f = 1/\hat{l} $), provided that the condition~\ref{eq:ang_diameter} is fulfilled.

Let us now consider how to separate the figure and scattering terms in HEW data. In absence of XRS, the mirror PSF would be independent of the energy and due only to figure errors (i.e. in the approximation of the geometrical optics). The resulting HEW would be also constant. Instead, due to the XRS, the figure PSF is convolved with the X-ray scattering PSF to return the PSF($\lambda$) being measured (Willingale~\cite{Willingale}; Stearns~et~al.~\cite{Stearns}; and many others),

\begin{equation}
	PSF(\lambda) = PSF_{\mathrm{fig}}\otimes PSF_{\mathrm{XRS}}(\lambda).
\label{eq:convolution}
\end{equation}

The resulting HEW will depend on the photon wavelength, as it does the PSF. In order to isolate the scattering term from the total PSF a deconvolution should be carried out, provided that the PSF$_{\mathrm{fig}}$ is known. However, if we assume that the XRS and the mirror deformations are statistically independent, the total HEW can be approximately calculated as the squared sum of the two contributions:

\begin{equation}
	HEW^2(\lambda) \approx HEW_{\mathrm{fig}}^2 + H^2(\lambda).
\label{eq:HEWsum}
\end{equation}

\noindent An estimation of HEW$_{\mathrm fig}$ can be obtained:
\begin{enumerate}
	\item{from the application of a ray-tracing code to several measurements of the mirror profile,}
	\item{from reliable extrapolation of the $HEW(\lambda)$ curve to $E \rightarrow 0$, in absence of low-energy diffraction effects like dust contamination, studied in detail by O'Dell~et~al. (\cite{ODell})}
	\item{from a direct measurement of the HEW in visible/UV light, provided that the diffraction at the mirror edges can be reliably calculated and subtracted.}
\end{enumerate} 

Once known the measured HEW$(\lambda)$ experimental trend and the HEW$_{\mathrm{fig}}$ term, the eq.~\ref{eq:HEWsum} can be used to isolate the scattering contribution from the experimental HEW trend: we shall prove in the next section that the $H(\lambda)$ function is immediately related to the reflecting surface {\it 1D Power Spectral Density} (PSD) $P(f)$

\begin{equation}
	P(f) = \frac{1}{L}\left|\int^L_0z(x)\mbox{e}^{-2\pi i f}\,\mbox{d}x\right|^2
\label{eq:PSDdef}
\end{equation}

\noindent where $z(x)$ is a height profile (of length $L$) of the mirror, measured in any direction (Stover~\cite{Stover}): the surface is assumed to be isotropic, and the spectral properties of the profile to be representative of the whole surface. The PSD is often measured in nm$^3$ units, and for optically-polished surfaces it is usually a decreasing function of the frequency $f$. 

PSD measurements have always a finite extent $[f_{\mathrm{min}}, f_{\mathrm{max}}]$, determined by the length and the spatial resolution of the measured profile. As well known, the surface rms $\sigma$ is simply computed from the PSD by integration over the spatial frequencies~$f$:
\begin{equation}
	\sigma^2 = \int_{f_{min}}^{f_{max}} P(f)\,\mbox{d}f
\label{eq:sigma}
\end{equation}
note that the integration range should always be specified.

\section{Estimation of $H(\lambda)$ for single-reflection focusing mirrors}\label{sect:fromPtoH}
\subsection{Single-layer coatings}\label{sect:singlelayer}
Firstly, we suppose the mirror to be plane and single-layer coated. For a surface with roughness rms $\sigma$, the specular beam intensity obeys the well-known {\it Debye-Waller} formula 
\begin{equation}
	R = R_{\mathrm F}\exp\left(-\frac{16\pi^2\sigma^2\sin^2\theta_{\mathrm i}}{\lambda^2}\right),
\label{eq:DW}
\end{equation}
\noindent here $R_F$ is the reflectivity at the grazing incidence angle $\theta_{\mathrm i}$, as calculated from Fresnel's equations (zero roughness). However, it should be noted in the eq.~\ref{eq:DW} that neither the spatial frequencies range where the PSD should be integrated is specified, nor the separation between reflected and scattered ray is clearly indicated: these ambiguities can be solved as follows.

Let us derive the {\it total scattered intensity} $I_{\mathrm s}$ from the conservation of the energy: for smooth surfaces, i.e. fulfilling the inequality $2\sigma \sin \theta_{\mathrm i} \ll \lambda$, we can approximate 
\begin{equation}
	I_{\mathrm s} = I_{\mathrm 0}R_{\mathrm F}\left[1-\exp\left(-\frac{16\pi^2\sigma^2\sin^2\theta_{\mathrm i}}{\lambda^2}\right)\right]\approx I_0R_F\frac{16\pi^2\sigma^2\sin^2\theta_{\mathrm i}}{\lambda^2}.
\label{eq:DWapprox}
\end{equation}
In grazing incidence, X-ray scattering lies mainly in the incidence plane. Moreover, the normalized scattered intensity per radian at the scattering angle $\theta_{\mathrm s}$ (either $\theta_{\mathrm s} > \theta_{\mathrm i}$ or $\theta_{\mathrm s} < \theta_{\mathrm i}$) is related to the PSD along with the well-known formula at first-order approximation (Church~et~al.~\cite{Church}; Church~\&~Tak\'{a}cs~\cite{ChurchTakacs}), valid for smooth, isotropic surfaces and for scattering directions close to the specular ray (i. e. $|\theta_{\mathrm s} - \theta_{\mathrm i}| \ll \theta_{\mathrm i}$),
\begin{equation}
	\frac{1}{I_{\mathrm 0}}\frac{\mbox{d}I_{\mathrm s}}{\mbox{d}\theta_{\mathrm s}} = \frac{16\pi^2}{\lambda^3}\sin^3\theta_{\mathrm i}R_{\mathrm F}P(f)
\label{eq:singlescatt}
\end{equation}
where $P(f)$ is the Power Spectral Density of the surface (eq.~\ref{eq:PSDdef}) and $I_{\mathrm 0}$ is the flux intensity of the incident X-rays. If the scattered intensity is evaluated at the scattering angle $\theta_{\mathrm s}$, the PSD can be immediately evaluated as a function of the spatial frequency $f$:
\begin{equation}
f = \hat{l}^{-1} = \frac{\cos\theta_{\mathrm i}-\cos\theta_{\mathrm s}}{\lambda} \approx \frac{\sin\theta_{\mathrm i}(\theta_{\mathrm s}-\theta_{\mathrm i})}{\lambda}.
	\label{eq:spacefreq}
\end{equation}
\noindent In the eq.~\ref{eq:spacefreq} the approximation was justified by the assumption $|\theta_{\mathrm s} - \theta_{\mathrm i}| \ll \theta_{\mathrm i}$ and the negative frequencies are conventionally assumed to scatter at $\theta_{\mathrm s} < \theta_{\mathrm i}$: the assumed approximations make the XRS diagram symmetric, because the PSD is an even function. 

For a single-reflection mirror shell, the extension of the formulae above-mentioned is straightforward by regarding $|\theta_{\mathrm s} - \theta_{\mathrm i}|$ as the angular distance at which the PSF is evaluated. The focal image is the superposition of many identical XRS diagrams on the image plane, generated by every meridional section of the mirror shell: since a $\pi$ angle rotation of every meridional plane of the shell sweeps the whole image plane, the scattered intensity is spread over a $\pi$ angle. The integration on circular coronae used to compute the mirror PSF (at positive angles) compensates this factor multiplying the XRS diagram by $2\pi$ (De~Korte~et~al.~\cite{DeKorte}). The remaining 2-fold factor accounts for the negative frequencies in the surface PSD. We shall henceforth suppose that the factor 2 is embedded in the PSD definition. Therefore, the eqs.~\ref{eq:singlescatt} and \ref{eq:spacefreq} can be used to describe the XRS contribution to the PSF.

We are now interested in the {\it scattered power at angles larger than a definite angle} $\alpha$ measured from the focus. Due to the steep fall of scattering intensity for increasing angles, the integral has a finite value
\begin{equation}
	I\left[|\theta_{\mathrm s} - \theta_{\mathrm i}|>\alpha\right] = \int_{\theta_{\mathrm i}+\alpha}^{\pi-\theta_i}\frac{\mbox{d}I_{\mathrm s}}{\mbox{d}\theta_{\mathrm s}}\,\mbox{d}\theta_{\mathrm s} .
\label{eq:intscatt1} 
\end{equation}
Combining eqs.~\ref{eq:singlescatt} and \ref{eq:intscatt1}, one obtains:
\begin{equation}
	I\left[|\theta_{\mathrm s} - \theta_{\mathrm i}|>\alpha\right] = I_{\mathrm 0}R_{\mathrm F}\frac{16\pi^2\sin^3\theta_{\mathrm i}}{\lambda^3}\int_{\theta_{\mathrm i}+\alpha}^{\pi-\theta_i}P(f)\,\mbox{d}\theta_{\mathrm s}
\label{eq:intscatt2}
\end{equation}
with respect to the definition used in the eqs.~\ref{eq:PSDdef} and \ref{eq:singlescatt}, a factor 2 was included in the PSD. The upper integration limit corresponds to a photon back-scattering: at first glance, this seems to violate our small-scattering angle assumption (eqs.~\ref{eq:singlescatt} and \ref{eq:spacefreq}), but it should be remembered that only the angles close to $\theta_{\mathrm i}$ contribute significantly to the integral in eq.~\ref{eq:intscatt1}: hence its value should not be significantly affected by a particular choice of the upper integration limit. After a variable change from $\theta_{\mathrm s}$ to $f$ (eq.~\ref{eq:spacefreq}), the eq.~\ref{eq:intscatt2} becomes (approximating $\cos\theta_{\mathrm i} \approx 1$ in the upper integration limit):
\begin{equation}
	I\left[|\theta_{\mathrm s} - \theta_{\mathrm i}|>\alpha\right] = I_{\mathrm 0}R_{\mathrm F}\frac{16\pi^2\sin^2\theta_{\mathrm i}}{\lambda^2}\int_{f_0}^{\frac{2}{\lambda}}P(f)\,\mbox{d}f
\label{eq:intscatt3}
\end{equation}
where $f_0 = \alpha \sin\theta_{\mathrm i} /\lambda$ is the spatial frequency corresponding to the scattering at the angle $\alpha$. As expected, this equation equals the integrated scattering according to the eq.~\ref{eq:DWapprox}, provided that we identify $I\left[|\theta_{\mathrm s} - \theta_{\mathrm i}|>\alpha\right]$ with $I_{\mathrm s}$, and the squared roughness rms with
\begin{equation}
	\sigma^2 =\int_{f_0}^{\frac{2}{\lambda}}P(f)\,\mbox{d}f.
\label{eq:sigmadef}
\end{equation}
The eq.~\ref{eq:sigmadef} is in agreement with the eq.~\ref{eq:sigma}, but it states clearly the window of spatial frequencies involved in the XRS. Therefore, for a definite angular limit $\alpha$ the "reflected beam" intensity can be simply calculated by using the Debye-Waller formula, provided that $\sigma^2$ is computed from the PSD integration beyond the frequency $f_{\mathrm 0}$, which corresponds to an X-ray scattering at $\alpha$. The upper integration limit is a very high frequency (close to $1/\mbox{\AA}$): hence, the atomic structure of the surface is not important in the integral of the eq.~\ref{eq:sigmadef}. Moreover, considering that the PSD trend for optically-polished surfaces decreases steeply for increasing $f$, the largest contribution to the integral should be given by the frequencies close to $f_0$.

Now we can evaluate $H(\lambda)$, the scattering term of the HEW. For simplicity, in the following we will suppose that the HEW is obtained from the collection of all the reflected/scattered photons: this allows us to avoid problems related to the finite size of the detector, and to extend the surface roughness PSD up to very large spatial frequencies. By definition, $H(\lambda)$ is twice the angular distance from focus at which the integrated scattered power halves the total reflected intensity:
\begin{equation}
	I\left[|\theta_{\mathrm s} - \theta_{\mathrm i}|>\alpha\right] =\frac{1}{2}I_{\mathrm 0}R_{\mathrm F}
\label{eq:HEWdef}
\end{equation}
we immediately derive, from the eq.~\ref{eq:DW},
\begin{equation}
	\exp\left(-\frac{16\pi^2\sigma^2\sin^2\theta_{\mathrm i}}{\lambda^2}\right) =\frac{1}{2},
\label{eq:DWdef}
\end{equation}
where $\sigma^2$ has now the meaning as per the eq.~\ref{eq:sigmadef}. Solving the eq.~\ref{eq:DWdef} for $\sigma^2$ and equating to the integral of the PSD,
\begin{equation}
	\int_{f_0}^{\frac{2}{\lambda}}P(f)\,\mbox{d}f = \frac{\lambda^2\ln 2}{16\pi^2\sin^2\theta_{\mathrm i}},
\label{eq:fstop}
\end{equation}
once known the PSD from topography measurements over a wide range of spatial frequencies, the PSD numerical integration in the eq.~\ref{eq:fstop} allows to recover $f_{\mathrm 0}$. In turn, $f_{\mathrm 0}$ is related to $H(\lambda)$ through the eq.~\ref{eq:spacefreq}, that we write in the following form
\begin{equation}
	H(\lambda) = \frac{2\lambda f_{\mathrm 0}}{\sin\theta_{\mathrm i}},
\label{eq:HEW}
\end{equation}
where $H$ is measured in radians. Note that the condition $H(\lambda)~\ll~\theta_{\mathrm i}$ is very important, for the eq.~\ref{eq:HEW} to hold. Small scattering angles and grazing incidence are also very important for the considerations that follow.   

\subsection{Multilayer coatings} \label{sect:multi}
The obtained result (eq.~\ref{eq:fstop}) can be extended to mirrors with {\it multilayer coatings}, used to enhance the grazing incidence reflectivity of mirrors in hard X-rays ($E > 10$ keV). In general, the multilayer cannot be characterized by means of a single PSD, due to the evolution of the roughness throughout the stack (Spiller~et~al.~\cite{Spiller93}; Stearns~et~al.~\cite{Stearns}). Moreover, due to the interference of scattered waves at each multilayer interface, the final scattering pattern is more structured than eq. \ref{eq:singlescatt}, with peaks whose height depends on the phase coherence of the interfaces (Kozhevnikov~\cite{Kozhevnikov}). The HEW term can be computed numerically from the XRS diagram.

In order to extend the eq.~\ref{eq:fstop} to mirrors coated with a {\it graded} multilayer, we have to assume the additional requirements:
\begin{enumerate}
	\item{the PSD is {\it constant} and completely {\it coherent} throughout the multilayer stack: i.e., the deposition process does not cause additional roughness and replicates simply the profile of the substrate. Therefore, all the PSDs and all the cross-correlation between interface profiles equal the PSD measured at the multilayer surface. This is often observed in the $\hat{l} > 10 \,\mu$m regime (Canestrari~et~al.~\cite{Canestrari}), where most of frequencies $f_{\mathrm 0}$ fall when the incidence angle is less than 0.5~deg. Most of microroughness growth, indeed, takes place for $ 10 \, \mu$m$\, > \hat{l} > 0.1 \, \mu$m. }
	\item{the multilayer reflectivity $R_{\lambda}(\theta_{\mathrm i})$ at the photon wavelength $\lambda$ changes gradually over angular scales of $H(\lambda)$. Ideally, this condition should be fulfilled by wideband multilayer coatings for astronomical X-ray mirrors.}
\end{enumerate}
 
Under these hypotheses, a quite tedious calculation reported in appendix \ref{sect:multilayers} shows that the eq.~\ref{eq:fstop} can be approximately applied also with multilayer coatings. The following developments also apply in that case.

\section{$H(\lambda)$ for a fractal surface}\label{sect:fractal}
We apply now the equations \ref{eq:fstop} and \ref{eq:HEW} to the typical (monodimensional) PSD model for optically-polished surfaces, {\it a power-law} (Church~\cite{Church88})
\begin{equation}
	P(f) = \frac{K_n}{f^n},
\label{eq:powerlaw}
\end{equation}
where the power-law index $n$ is a {\it real} number in the interval $1<n<3$ and $K_n$ is a normalization factor. A power-law PSD is typical of a {\it fractal surface}, and it represents the high-frequency regime of a K-correlation model PSD (Stover~\cite{Stover}). This model exhibits a saturation for $f \rightarrow~0$ that avoids the PSD divergence. In practice, the fractal behavior dominates in almost all spatial frequencies of interest for X-ray optics. 

\begin{figure}
	\resizebox{\hsize}{!}{\includegraphics{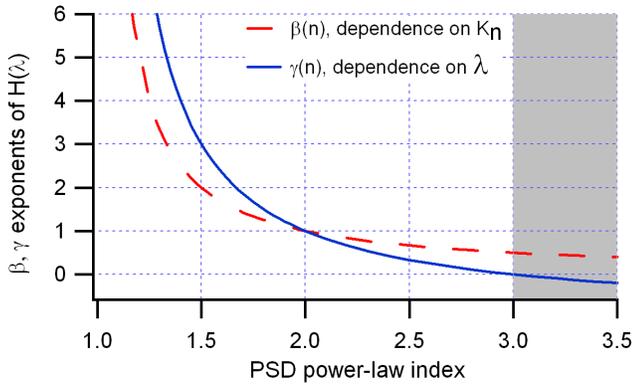}}
	\caption{Dependence of the spectral exponents for different indexes $n$ of a power-law PSD, for a single-reflection focusing mirror. In the forbidden region ($ n > 3$) $\gamma$ would be negative.}
\label{fig:beta_gamma}
\end{figure}

There are interesting reasons for which $n$ can take values on the interval (1:3). In fact, for a surface in the 3D space, $n$ is related to its {\it Hausdorff-Besicovitch dimension} $D$ (Barab\'{a}si~\&~Stanley~\cite{BarabasiStanley}) along with the equation $n=7-2D$ (see Church~\cite{Church88}; Gouyet~\cite{Gouyet}). The restriction $1<n<3$ for a fractal surface is therefore necessary to have $3>D>2$.

A power-law PSD is particularly interesting because the integral on left-hand side of the eq.~\ref{eq:fstop} can be explicitly calculated:
\begin{equation}
	K_n \frac{f_{\mathrm 0}^{1-n}-\left(\frac{2}{\lambda}\right)^{1-n}}{n-1}=\frac{\lambda^2\ln2}{16\pi^2\sin^2\theta_{\mathrm i}}.
\label{eq:powerlaw_fstop}
\end{equation}
As $1-n < 0$, in grazing incidence the $(2/\lambda)^{1-n}$ term can be neglected with respect to $f_{\mathrm 0}^{1-n}$. By isolating the frequency $f_{\mathrm 0}$ and using the eq.~\ref{eq:HEW} to derive $H(\lambda)$, we obtain after some algebra, for the scattering term of the HEW, 
\begin{equation}
	H(\lambda) = 2\left[\frac{16\pi^2 K_n}{(n-1)\ln 2}\right]^{\frac{1}{n-1}}\left(\frac{\sin\theta_{\mathrm i}}{\lambda}\right)^{\frac{3-n}{n-1}}.
\label{eq:HEW_powerlaw}
\end{equation}
This equation states that:

\begin{enumerate}
\item{The $H(\lambda)$ function for a power-law PSD has a power-law dependence on the photon energy $E \propto 1/\lambda$, i.e., $ H(E)~\propto~E^{\gamma}$. The power-law index $\gamma$ is related to the PSD power-law index $n$ through the simple equation:}
\begin{equation}
	\gamma = \frac{3-n}{n-1}.
\label{eq:gamma_index}
\end{equation}

\noindent As $1<n<3$, $\gamma$ is positive, i.e. $H$ is an increasing function of the photon energy. For a fixed value of $K_n$, the HEW diverges quickly for $n \approx 1$ but very slowly for $n \approx 3$: a PSD power-law index close to 2-3 would hence be preferable in order to reduce the degradation of focusing performances for increasing energies.

\item{$H(\lambda)$ depends on the sine of the incidence angle at the $\gamma^{\mathrm t\mathrm h}$ power. In other words, the HEW depends only on the ratio $\sin\theta_{\mathrm i} /\lambda$: this {\it scaling relation} shows that for a given power-law PSD (with $n<3$) at a given photon wavelength $\lambda$ we can reduce the HEW by decreasing the incidence angle.}
\item{$H(\lambda)$ increases with the PSD normalization $K_n$, as expected: the dependence is also a power law with spectral index 
\begin{equation}
	\beta = \frac{1}{n-1}.
\label{eq:beta_index}
\end{equation}
As for $\gamma(n)$, the closeness of $n$ to the maximum allowed value for fractal surfaces makes less severe the roughness effect on imaging degradation.}
\end{enumerate}

\begin{figure}
	\resizebox{\hsize}{!}{\includegraphics{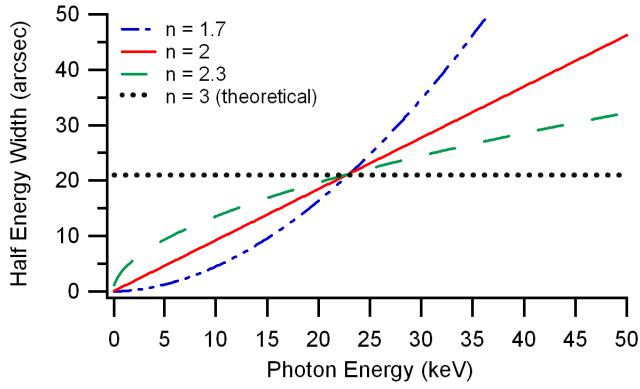}}
	\caption{$H(E)$ simulations assuming power-law PSDs with constant $\sigma$~=~4~$\mbox{\AA}$ in the spatial wavelengths range [$100 \div 0.01 \mathrm\mu m]$, but variable power-law index $n$. The incidence angle is fixed at $\theta_{\mathrm i}$ = 0.5 deg.}
\label{fig:powersim1}
\end{figure}

The functions $\beta$ and $\gamma$ are plotted in fig.~\ref{fig:beta_gamma}. For instance, if $n=~2$, $\gamma\!=\!\beta\!=\!1$, and $H(E)$ increases linearly with both photon energy and $K_n$ coefficient. The divergence of indexes $\beta,\gamma$ for $n~\approx~1$ makes apparent the importance of obtaining steep PSDs in the optical polishing of X-ray mirrors. Finally, it is worth noting that for $n > 3$ there is the {\it theoretical} possibility of a slight decrease of $H(E)$ for increasing energy because $\gamma(n)$ becomes {\it negative}. 

\begin{figure}
	\resizebox{\hsize}{!}{\includegraphics{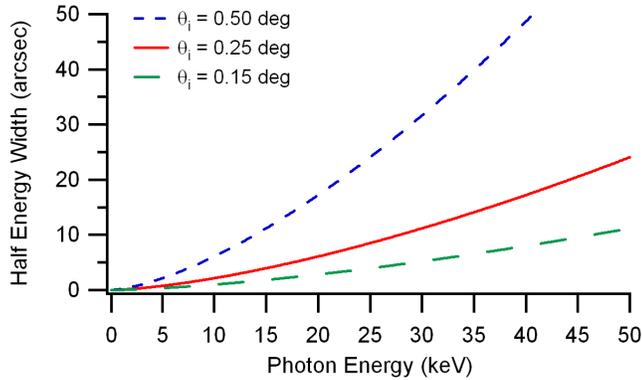}}
	\caption{$H(E)$ simulations assuming a power-law PSD with power-law index $n = 1.8$ and with $\sigma$~=~4~$\mbox{\AA}$ in the spatial wavelengths range [100 $\div$ 0.01 $\mu$m], but variable incidence angle $\theta_{\mathrm i}$.}
\label{fig:powersim2}
\end{figure}

To clarify the dependence of the HEW on the power-law index $n$ and the incidence angle, we depict in fig.~\ref{fig:powersim1} and \ref{fig:powersim2} some examples of $H(E)$ simulations (single reflection) for some power-law PSDs in the photon energy range 0.1-50~keV. The $H(E)$ curves were computed using the eq.~\ref{eq:HEW_powerlaw}. In fig.~\ref{fig:powersim1} the incidence angle $\theta_{\mathrm i}$ is fixed at 0.5 deg and the index $n$ is variable; a constant $n = 1.8$ and a variable $\theta_{\mathrm i}$ is instead assumed in the simulations of fig.~\ref{fig:powersim2}. Note in fig.~\ref{fig:powersim1} the slower $H(E)$ increase for larger $n$ and the common intersection point, determined by the particular choice of the incidence angle and the $\sigma$ = 4 $\mbox{\AA}$ value in the window of spatial wavelengths [100 $\div$ 0.01 $\mu$m]. 

\section{Numerical integration of the PSD}\label{sect:numerical}
A power-law PSD is a modelization that can be used for optically-polished surfaces. If the polishing process is not optimized or a reflecting layer is grown onto a optically polished substrate, several deviations from a power-law trend can be observed. A typical "bump", for instance, can be present in the PSD of multilayer coatings, often in the range of spatial wavelengths $[10 \, \div \, 0.1 \, \mu {\mathrm m}]$, as a result of the replication of the substrate topography and of fluctuations intrinsically related to the random deposition process (Spiller~et~al.~\cite{Spiller93}; Stearns~et~al.~\cite{Stearns}). If the PSD deviates significantly from a power-law, the eq.~\ref{eq:HEW_powerlaw} cannot be used. However, if the surface PSD has been extensively measured over a wide range of spatial frequencies $[f_{\mathrm m}, f_{\mathrm M}]$ (wide enough to have $f_{\mathrm m} < f_{\mathrm 0}(\lambda)$ for all $\lambda$), the HEW scattering term $H(\lambda)$ can be computed by numerical integration (eqs. \ref{eq:fstop} and \ref{eq:HEW}), on condition that the following approximation is valid:  
\begin{equation}
	\int^{\frac{2}{\lambda}}_{f_0}P(f)\,\mbox{d}f \approx \int^{f_M}_{f_0}P(f)\,\mbox{d}f .
\label{eq:integrapprox}
\end{equation}
The condition above is usually satisfied when $f_{\mathrm 0} \ll f_{\mathrm M} $ i.e. when the following inequality holds:
\begin{equation}
	H(\lambda) \ll \frac{2\lambda f_{\mathrm M}}{\sin\theta_{\mathrm i}}.
\label{eq:HEWapprox}
\end{equation}

\noindent As we are also interested in computing $H(\lambda)$ in hard X-rays (small $\lambda$), there is the possibility that the two integrals in the eq.~\ref{eq:integrapprox} differ by a significant factor. In this case the integral can be corrected by adding the remaining term
\begin{equation}
	\int^{\frac{2}{\lambda}}_{f_0}P(f)\,\mbox{d}f = \int^{f_M}_{f_0}P(f)\,\mbox{d}f +\int^{\frac{2}{\lambda}}_{f_M}P(f)\,\mbox{d}f ,
\label{eq:integrcorr}
\end{equation}
that can be evaluated, in principle, by measuring the mirror reflectivity within an angular acceptance corresponding to the spatial frequency $f_{\mathrm M}$, and using the Debye-Waller formula to derive $\sigma^2$; then, the importance of measuring the PSD in a {\it very} wide frequencies interval becomes apparent. The value of $f_{\mathrm 0}$ depends strongly on both incidence angle and photon energy: for soft X-rays ($<$ 10 keV) and very small angles ($<$ 0.2 deg) the characteristic spatial wavelength $\hat{l} = 1/f_{\mathrm 0}$ often falls in the millimeter or centimeter range. 

	It should be noted that, if the detector is small, a fraction of the scattered photons can be lost; to account for the finite angular radius of the detector $d$ (as seen from the optic principal plane), one should integrate the PSD over the smaller interval [$f_{\mathrm 0}$,~$d \sin\theta_{\mathrm i}/\lambda$] to recover the measured $H(\lambda)$ trend. As an alternative method, one can compare the theoretical predictions of eqs.~\ref{eq:fstop} and \ref{eq:HEW} with the experimental $H(\lambda)$ values, as calculated from the Encircled Energy normalized to the photon count foreseen by the Fresnel equations (i.e. with zero roughness), rather than to the maximum of the measured Encircled Energy function.

\section{Computation of the PSD from the $H(\lambda)$ trend}\label{sect:fromHtoPSD}

If the approach described above can be used to simulate the HEW trend from a measured surface PSD, the reverse problem, i.e. the derivation of surface PSD from the measured HEW trend is also possible. This requires that {\it the figure error contribution had been reliably measured}, in order to isolate the scattering term function $H(\lambda)$ using the eq.~\ref{eq:HEWsum}.

This problem is interesting for three reasons at least: 
\begin{enumerate}
	\item{it is a quick, non-destructive surface characterization method in terms of its PSD.}
	\item{The measurement is extended to a large portion of the illuminated optic, hence local surface features are averaged and ruled out from the PSD.}
	\item{For a given HEW($\lambda$) requirement in the telescope sensitivity energy band, it allows establishing the maximum allowed PSD.}
\end{enumerate}

In order to find an analytical expression for the PSD, we note that the spatial frequency $f_{\mathrm 0}$ that scatters at an angular distance $H/2$ from the specular beam is a function only of $\lambda$, along with the eq.~\ref{eq:HEW}. Solving for $f_{\mathrm 0}$, we have
\begin{equation}
	f_{\mathrm 0}(\lambda) \approx H(\lambda)\frac{\sin\theta_{\mathrm i}}{2\lambda}.
\label{eq:fstopfromH}
\end{equation}
We suppose that all scattered photons are collected, so we can assume the eq.~\ref{eq:fstop} as valid. By deriving both sides of eq.~\ref{eq:fstop} with respect to $\lambda$, we have
\begin{equation}
	\frac{\mbox{d}}{\mbox{d}\lambda}\left(\int^{\frac{2}{\lambda}}_{f_0}P(f)\,\mbox{d}f \right)=\frac{\ln 2}{8\pi^2\sin^2\theta_{\mathrm i}}\lambda,
\label{eq:derivefstop1}
\end{equation}
that is,
\begin{equation}
	-\frac{2}{\lambda^2} P\left(\frac{2}{\lambda}\right)-\frac{\mbox{d}f_{\mathrm 0}}{\mbox{d}\lambda} P(f_{\mathrm 0})=\frac{\ln 2}{8\pi^2\sin^2\theta_{\mathrm i}}\lambda,
\label{eq:derivefstop2}
\end{equation}
and, using the eq.~\ref{eq:fstopfromH} to compute the derivative of $f_{\mathrm 0}$:
\begin{equation}
	\frac{-\frac{2}{\lambda}P\left(\frac{2}{\lambda}\right)+f_{\mathrm 0} P(f_{\mathrm 0})}{\lambda} -\frac{\sin\theta_{\mathrm i}}{2\lambda}\frac{\mbox{d}H(\lambda)}{\mbox{d}\lambda} P(f_{\mathrm 0})=\frac{\ln 2}{8\pi^2\sin^2\theta_{\mathrm i}}\lambda.
\label{eq:derivefstop3}
\end{equation}
Now remember that, in grazing incidence, $f_{\mathrm 0} \ll 2\lambda^{-1}$ by several orders of magnitude. Even if $P(f)$ is not a power-law, it is always a steeply decreasing function of $f$. Moreover, it should have over [$f_{\mathrm 0}, 2\lambda^{-1}$] an average PSD index $\tilde{n}>1$, for the reasons explained in the sect.~\ref{sect:fractal}. This means that
\begin{equation}
	\frac{P(f_0)}{P(2\lambda^{-1})} \approx \left(\frac{2\lambda^{-1}}{f_0}\right)^{\tilde{n}} \gg \frac{2\lambda^{-1}}{f_0},
\label{eq:derivefstop4}
\end{equation}
therefore, in practical cases the $2\lambda^{-1}P(2\lambda^{-1})$ term in the eq.~\ref{eq:derivefstop3} is negligible with respect to $f_{\mathrm 0}P(f_{\mathrm 0})$. Consequently, we can neglect the first term of eq.~\ref{eq:derivefstop2}: then we have 
\begin{equation}
	-\frac{\mbox{d}f_{\mathrm 0}}{\mbox{d}\lambda} P(f_{\mathrm 0})\approx \frac{\ln 2}{8\pi^2\sin^2\theta_{\mathrm i}}\lambda.
\label{eq:derivefstop5}
\end{equation}
Combining this with the eq.~\ref{eq:fstopfromH} and collecting the constants, we obtain the final result
\begin{equation}
	\frac{P(f_{\mathrm 0})}{\lambda}\,\frac{\mbox{d}}{\mbox{d}\lambda}\!\left(\frac{H(\lambda)}{\lambda}\right) +\frac{\ln 2}{4\pi^2\sin^3\theta_{\mathrm i}} \approx 0.
\label{eq:final}
\end{equation}

The eq.~\ref{eq:final} {\it enables the computation of the PSD} \,(at the spatial frequency given by the eq.~\ref{eq:fstopfromH}) {\it along with the derivative of the ratio $H(\lambda)/\lambda $ with respect to $\lambda$}.
 
	The obtained equation shows that $P(f)$ is {\it inversely proportional} to the derivative of $H(\lambda)/\lambda$. This result seems strange at first glance, because by decreasing $H(\lambda)$ one would obtain a larger $P(f)$ (a rougher surface). One should remember, indeed, that by reducing $H(\lambda)$ we increase $P(f_{\mathrm 0})$, but $f_{\mathrm 0}$ is shifted towards the low frequencies domain, where $P(f_{\mathrm 0})$ is expected to be higher. In fact, the "rough" or "smooth" feature of the surface depends on whether $f_{\mathrm 0}$ or $P(f_{\mathrm 0})$ varies more rapidly, i.e. on the overall $H(\lambda)$ trend.

	We can also check the correctness of the eq.~\ref{eq:final} by computing the PSD for the particular case of the $H(\lambda)$ derived from the integration of a power-law PSD (the eq.~\ref{eq:HEW_powerlaw}, derived under the same approximation, the eq.~\ref{eq:derivefstop4}). If the results are correct, the substitution of the HEW trend of the eq.~\ref{eq:HEW_powerlaw} in the eq.~\ref{eq:final} should return the original PSD (eq.~\ref{eq:powerlaw}). The straightforward, but lengthy calculation (carried out in appendix~\ref{sect:powerlawcalc}) shows that the substitution returns
\begin{equation}
	P(f_0) = \frac{K_n}{f_{\mathrm 0}^n}
\label{eq:powerlaw2}
\end{equation}
as expected.

The eq.~\ref{eq:final} should be approximately valid also for graded multilayers with a slowly-decreasing reflectivity (see sect.~\ref{sect:multi}), however, due to the approximations needed to extend the eq.~\ref{eq:fstop} to the multilayers, the resulting PSD should be considered a "first guess" in this case. Then, the matching of the PSD to the required HEW trend should be checked by means of a detailed computation of the XRS PSF($\lambda$).

\section{Extension to X-ray mirrors with multiple reflections}\label{sect:multiple}

The formalism exposed in the previous sections can be extended to a double-reflection optic (like a Wolter-I one). In this optical configuration, photons are firstly reflected by a parabolic surface and subsequently by a hyperbolic one. If the smooth-surface condition is satisfied, multiple scattering is often negligible (Willingale~\cite{Willingale}) and the scattering diagrams of the two reflecting surfaces can be simply summed (De~Korte~et~al.~\cite{DeKorte}; Stearns~et~al.~\cite{Stearns}). The source is assumed to be at infinite distance, then X-rays impinge on the two surfaces at the same angle $\theta_{\mathrm i}$. If the surface PSDs are the same for both reflections, the scattering diagram will be simply doubled. Thus, the integrated scattered intensity is also doubled: 
\begin{equation}
	I_{\mathrm s} = 2I_{\mathrm 0} R^2_{\mathrm F}\left[1-\exp\left(-\frac{16\pi^2\sigma^2\sin^2\theta_{\mathrm i}}{\lambda^2}\right)\right].
	\label{eq:2refl}
\end{equation}
The $R_{\mathrm F}$ factor is squared in the eq.~\ref{eq:2refl} because each ray is reflected {\it twice}: in absence of scattering the reflected power would be $I_{\mathrm 0} R^2_{\mathrm F}$, so the half-power scattering angle condition reads 
\begin{equation}
	I\left[|\theta_{\mathrm s} - \theta_{\mathrm i}|>\frac{H(\lambda)}{2}\right] =\frac{1}{2}I_{\mathrm 0}R^2_{\mathrm F},
	\label{eq:2HEWdef}
\end{equation}
and, combining the eqs.~\ref{eq:2refl} and \ref{eq:2HEWdef}, we obtain
\begin{equation}
	\exp\left(-\frac{16\pi^2\sigma^2\sin^2\theta_{\mathrm i}}{\lambda^2}\right) = \frac{3}{4}.
	\label{eq:2exp}
\end{equation}
Solving for $\sigma^2$, and using the eq. \ref{eq:sigmadef},
\begin{equation}
	\int_{f_0}^{\frac{2}{\lambda}}P(f)\,\mbox{d}f = \frac{\lambda^2\ln(4/3)}{16\pi^2\sin^2\theta_{\mathrm i}},
	\label{eq:2fstop}
\end{equation}
that differs from the eq. \ref{eq:fstop} only in the factor $\ln(4/3)$ instead of $\ln 2$ on right-hand side. Consequently, the corresponding differential equation is
\begin{equation}
	\frac{P(f_{\mathrm 0})}{\lambda}\,\frac{\mbox{d}}{\mbox{d}\lambda}\!\left(\frac{H(\lambda)}{\lambda}\right) +\frac{\ln (4/3)}{4\pi^2\sin^3\theta_{\mathrm i}} \approx 0.
	\label{eq:2final}
\end{equation}

Similar equations can be derived for an optical system with an arbitrary number of reflections $N$: to compute the $H(\lambda)$ from the PSD, 
\begin{equation}
	\int_{f_0}^{\frac{2}{\lambda}}P(f)\,\mbox{d}f = \frac{\lambda^2}{16\pi^2\sin^2\theta_{\mathrm i}}\ln\left(\frac{2N}{2N-1}\right).
	\label{eq:Nfstop}
\end{equation}
If the PSD is a power-law $P(f) = K_n/f^ n $ we can generalize the eq. \ref{eq:HEW_powerlaw}:
\begin{equation}
	H(\lambda) =2\left[\ln\left(\frac{2N}{2N-1}\right)\right]^{\frac{1}{1-n}}\left[\frac{16\pi^2 K_n}{(n-1)}\right]^{\frac{1}{n-1}}\left(\frac{\sin\theta_{\mathrm i}}{\lambda}\right)^{\frac{3-n}{n-1}},
\label{eq:NHEW_powerlaw}
\end{equation}
note the divergence of the logaritmic factor for increasing $N$, due to the negative exponent $1/(1-n)$. This indicates that $H(\lambda)$ increases rapidly with the number of reflections, as expected. 

Finally, we can also generalize the differential eq.~\ref{eq:final} to an arbitrary number of reflections,
\begin{equation}
	\frac{P(f_{\mathrm 0})}{\lambda}\,\frac{\mbox{d}}{\mbox{d}\lambda}\!\left(\frac{H(\lambda)}{\lambda}\right) +\frac{\ln \left(\frac{2N}{2N-1}\right)}{4\pi^2\sin^3\theta_{\mathrm i}} \approx 0.
	\label{eq:Nfinal}
\end{equation}
In the eqs.~\ref{eq:Nfstop} and \ref{eq:Nfinal}, $f_{\mathrm 0}$ is always related to $ H(\lambda)$ by the eq.~\ref{eq:fstopfromH}.

\section{An example}\label{sect:example}

\begin{figure}
	\resizebox{\hsize}{!}{\includegraphics{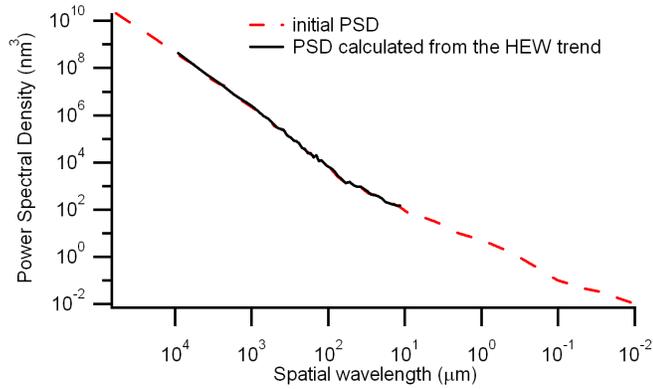}}
	\caption{an hypothetical PSD with reasonable values and a PSD break around a 100 $\mu$m spatial wavelength (dashed line). This PSD is adopted to compute the corresponding HEW trends for 1,2,3 reflections at a 0.3~deg grazing incidence angle (fig.~\ref{fig:HEW_exp}). The achieved HEW trends were used to re-calculate the respective PSDs (solid line). For clarity, we do not plot the single PSDs, but just their overlap.}
\label{fig:PSD_exp}
\end{figure}

As an application of the equations reported above, we shall make use of a {\it simulated} surface PSD with reasonable values, that is {\it not} a power-law. The PSD (see fig.~\ref{fig:PSD_exp}, dashed line) is extended from $10^5 \, \mu$m down to a $0.01 \, \mu$m spatial wavelength, with a break around $100 \, \mu$m: at the lowest frequencies the PSD is steep ($n \approx 2.3$), whereas at the largest frequencies it is smoother ($n \approx 1.3$). From the discussion in sect.~\ref{sect:fractal} concerning the relation between the exponents of the PSD and the HEW (eq. \ref{eq:gamma_index}), we should expect that the PSD break causes a slope change in the function $H(\lambda)$: however, as the actual PSD is not a power-law, the $H(\lambda)$ function should be computed by means of the eqs.~\ref{eq:HEW} and \ref{eq:Nfstop}. Before carrying out the integration, we can remark qualitatively that, as we increase the photon energy, the highest frequencies in the PSD (where the PSD index becomes smaller) become important; hence, we can expect a steeper increase of the HEW at the highest energies.

\begin{figure}
	\resizebox{\hsize}{!}{\includegraphics{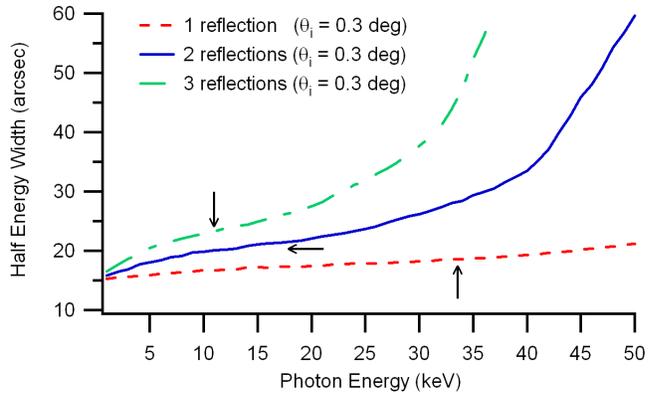}}
	\caption{the HEW trend computed from the PSD for 1,2,3 reflections, plus 15 arcsec of HEW due to figure errors. The HEW trends were used to compute back the PSD (the solid line in fig.~\ref{fig:PSD_exp}) to verify the reversibility of the calculation. The energy at which the concavity change takes place is also indicated (arrows).}
\label{fig:HEW_exp}
\end{figure}

The analysis is made quantitative in fig.~\ref{fig:HEW_exp}, where we show the calculated HEW trends from the PSD in fig.~\ref{fig:PSD_exp} (the dashed line) by means of the eqs.~\ref{eq:HEW} and \ref{eq:Nfstop}, assuming 1,2,3 reflections at the same grazing incidence angle (0.3 deg). The approximation of eq.~\ref{eq:integrapprox} was adopted. In addition to the scattering term, 15 arcsec of HEW due to figure errors were added in quadrature. The HEW increases slowly (concave downwards) at low energies, corresponding to a frequency $f_{\mathrm 0}$ in the steeper part of the PSD. Then it increases more steeply (concave upwards) when the energy becomes large enough to set $f_{\mathrm 0}$ in the portion of the spectrum with $n \approx 1.3$. By increasing the number of reflections, the HEW values also increase, and the "turning point" where the HEW starts to diverge (arrows in fig.~\ref{fig:HEW_exp}) shifts at lower X-ray energies. All the calculation is based on the assumption that the contribution of the PSD over the maximum measured frequency $f_{\mathrm M} = 0.01 \,\mu$m is negligible. Otherwise, the computed HEW values will be underestimated (see sect.~\ref{sect:numerical}).

In addition to the general trend of the HEW, there are oscillations due to small irregularities in the adopted PSD: the calculation is, in fact, very sensitive to small variations of the PSD values. Notice that for a definite energy {\it all} the frequencies larger than $f_{\mathrm 0}$ contribute to the HEW value, even if the largest contribution comes from frequencies near $f_{\mathrm 0}$: this is a consequence of the steeply decreasing trend of the PSD. 

We checked the reversibility of the result by computing the PSD from the HEW trends (after subtracting in quadrature 15 arcsec figure error) by means of the eq.~\ref{eq:Nfinal} with the respective value of $N$. The resulting PSDs (the solid line in fig.~\ref{fig:PSD_exp}) were overplotted to the initial PSD, with a perfect superposition. Each obtained PSD has, indeed, an extent of spatial frequencies smaller than the initial one: the overall PSD ranges from $10^4$ to $11 \,\mu$m (vs. the initial $10^5 \,\div \,0.01 \, \mu$m), and the smaller wavelengths could be computed from the HEW trend with $N=3$. The limitation in spatial frequency ranges occurs for two reasons:
\begin{enumerate}
\item{{\it small f $-$ large $\hat{l}$}: all the power scattered by the {\it lowest frequencies} is found at angles less than 1/2 HEW even for the lowest energies being considered: therefore, that part of the spectrum is not necessary to compute the HEW in the energy range of interest;} 
\item{{\it large f $-$ small $\hat{l}$}: the PSD is computed from a derivative, therefore the information concerning the absolute magnitude of the HEW is substantially lost. This information was included in the integral of the PSD (eq.~\ref{eq:Nfstop}) for the maximum considered energy.}
\end{enumerate}

Therefore, from the integral in the eq.~\ref{eq:Nfstop} we cannot recover the PSD over the minimum computed spatial wavelength ($11 \, \mu$m, using the HEW trend with $N=3$), but we can at least calculate the value of $\sigma$ at spatial wavelengths smaller than $11 \, \mu$m. Substituting the incidence angle and the minimum photon wavelength being considered ($\lambda = 0.24 \,\mbox{\AA}$) in the eq.~\ref{eq:Nfstop} with $N =3$ and with the approximation of the eq.~\ref{eq:integrapprox}, we obtain $\sigma = 1.6 \,\mbox{\AA}$, in perfect agreement with the value computed from the original PSD. 

Summing up, for a given incidence angle the $H(\lambda)$ function in a definite photon energy range is equivalent to the PSD in a corresponding range of spatial frequencies $f$ (or equivalently, spatial wavelengths $\hat{l}$), {\it plus} the integral of the PSD beyond the maximum frequency being computed. Therefore, requirements of a definite HEW($\lambda$) function in designing an X-ray optical system can be translated in terms of PSD in a frequencies range [$f_{\mathrm{min}}, f_{\mathrm{max}}$] plus the surface rms at frequencies beyond $f_{\mathrm{max}}$. The usefulness of such a relationship is apparent.

\section{Conclusions}
In the previous pages we have developed useful equations to compute the contribution of the X-ray scattering to the HEW of a grazing incidence X-ray optic, by means of a simple integration. The formalism has been inverted in order to derive the PSD of the surface from the function $H(\lambda)$, and it can be extended to an arbitrary number of reflections at the same incidence angle. The equations are valid for a single-layer coating mirror, but they can be approximately applied to a multilayer-coated mirror. This approach is particularly useful in order to establish the surface finishing level needed to keep the X-ray scattering HEW of X-ray optics within the limits fixed by the X-ray telescope requirements.

It should be remarked that the reasoning was developed for the Half-Energy Width, but it can be extended to {\it any} angular diameter including a fraction $\eta$ of the energy spread around the focal point. To do this, it is sufficient to substitute the logarithmic factors in equations \ref{eq:Nfstop}, \ref{eq:NHEW_powerlaw}, \ref{eq:Nfinal},

\begin{equation}
	\ln\left(\frac{2N}{2N-1}\right)\rightarrow \ln\left(\frac{N}{N-1+\eta}\right),
	\label{eq:logfact}
\end{equation}
and for instance, to compute the 90\%-energy diameter for a double reflection mirror, simply substitute $\eta = 0.9$ and $N =2$. The proof is straightforward: however, one should always keep in mind that the energy diameters computed with this method can be considered valid only if they are much smaller than the incidence angle $\theta_{\mathrm i}$.

Notice that in the development of the exposed formalism we have supposed, in addition to the smooth-surface condition, two additional hypotheses:
\begin{enumerate}
\item{the source is at infinite distance from the mirror}
\item{the X-ray detector is large enough to collect {\it all} the scattered photons.}
\end{enumerate}
 In order to apply the mentioned equations to experimental calibrations of X-ray optics at existing facilities (like MPE-PANTER), where the source is at a finite distance and the detector has a finite size, some corrections should be taken into account. We will deal with their quantification in a subsequent paper.

\appendix

\section{Extension to multilayer coatings}\label{sect:multilayers}
Here we provide with a plausibility argument to extend the formalism of sect.~\ref{sect:singlelayer} to mirror shells with multilayer coatings (see sect.~\ref{sect:multi}). The intensity of a scattered wave at each interface is proportional to its PSD as per the eq.~\ref{eq:singlescatt}, and the overall scattering diagram will be their {\it coherent} interference. To simplify the notation, we neglect the X-ray refraction and we suppose that the incidence angle is beyond the critical angles of the multilayer components. The electric field scattered by the $k^{\mathrm t\mathrm h}$ interface can be written as
\begin{equation}
 	E_k = E_{\mathrm 0} T_k r_k X_k(f) \exp(-{\mathrm i}\phi_k),
	\label{eq:pass_1}
\end{equation}
where $r_k$ is the single-boundary amplitude reflectivity, $E_{\mathrm 0}$ the incident electric field {\it amplitude}, the weights $T_k$ are the relative amplitudes of the electric field in the stack (in scalar, single-scattering approximation), and account for the extinction of the incident X-rays due to gradual reflection and absorption. $X_k(f)$ is the single-boundary scattering power (proportional to the $PSD(f)$ amplitude), and $\phi_k$ is the phase of the scattered wave at $\theta_{\mathrm s}$ by the $k^{\mathrm t\mathrm h}$ interface
\begin{equation}
	\phi_k = 2\pi\frac{\sin\theta_{\mathrm i} +\sin\theta_{\mathrm s}}{\lambda}z_k,
 	\label{eq:pass_2}
\end{equation}
where $z_k$ is the depth of the $k^{\mathrm t\mathrm h}$ interface with respect to the outer surface of the multilayer. Now, the measured intensity is 
\begin{equation}
	|E_{\mathrm scatt}|^2 = \left|\sum^N_{k=0}E_k\right|^2 = |E_{\mathrm 0}|^2|X_k(f)|^2\left|\sum^N_{k=0}r_k T_k \exp(-{\mathrm i}\phi_k)\right|^2.
 	\label{eq:pass_3}
\end{equation}
Now, $|E_0|^2 = I_0$, the incident X-ray flux intensity, and $|X_k(f)|^2$ is proportional to the interfacial PSD $P(f)$, which is independent of $k$ by hypothesis. Assuming the proportionality factor of eq.~\ref{eq:singlescatt} for $|X_k(f)|^2$, we obtain for the scattering diagram
\begin{equation}
	\frac{1}{I_{\mathrm 0}}\frac{\mbox{d}I_{\mathrm s}}{\mbox{d}\theta_{\mathrm s}} =\frac{16\pi^2}{\lambda^3}\sin^3\theta_{\mathrm i}P(f)\left|\sum^N_{k=0}r_k T_k \exp(-{\mathrm i}\phi_k)\right|^2
	\label{eq:pass_4}
\end{equation}
and if we set
\begin{equation}
	K_{\lambda}(\theta_{\mathrm i}, \theta_{\mathrm s})= \left|\sum^N_{k=0}r_k T_k \exp(-{\mathrm i}\phi_k)\right|^2,
	\label{eq:pass_5}
\end{equation}
the eq.~\ref{eq:pass_4} becomes analogous to the eq.~\ref{eq:singlescatt}, with $K_{\lambda}(\theta_{\mathrm i}, \theta_{\mathrm s})$ playing the role of $R_{\mathrm F}$. Note that $K_{\lambda}(\theta_{\mathrm i}, \theta_{\mathrm i}) = R_{\lambda}(\theta_{\mathrm i})$, the multilayer reflectivity in single reflection approximation. As before, we write the scattering diagram for a mirror with axial symmetry as a function of the angular distance from the focus $\alpha = |\theta_{\mathrm i} - \theta_{\mathrm s}|$ averaging the contributions of negative and positive frequencies
\begin{equation}
	\frac{1}{I_{\mathrm 0}}\frac{\mbox{d}I_{\mathrm s}}{\mbox{d}\theta_{\mathrm s}} =\frac{8\pi^2}{\lambda^3}\sin^3\theta_{\mathrm i}P(f)[K_{\lambda}(\theta_{\mathrm i}, \theta_{\mathrm i}-\alpha)+K_{\lambda}(\theta_{\mathrm i}, \theta_{\mathrm i}+\alpha)].
	\label{eq:pass_6}
\end{equation}
For a single reflection optic, we can calculate the scattered power over $H/2$, where $H$ is the scattering term of optic Half-Energy Width:
\begin{equation}
  	I_{\mathrm s}[\alpha > H/2] = \frac{1}{2}I_{\mathrm 0}R_{\lambda}(\theta_{\mathrm i}).
	\label{eq:pass_7}
\end{equation}
Now, the steps \ref{eq:DWdef} and \ref{eq:fstop} can be repeated:
\begin{equation}
	\int_{f_0}^{\frac{2}{\lambda}}\frac{K_{\lambda}(\theta_{\mathrm i},\theta_{\mathrm i}+\alpha)+K_{\lambda}(\theta_{\mathrm i},\theta_{\mathrm i}-\alpha)}{R_{\lambda}(\theta_{\mathrm i})}P(f)\,\mbox{d}f =\frac{\lambda^2\ln 2}{8\pi^2\sin^2\theta_{\mathrm i}}
	\label{eq:pass_8}
\end{equation}
where $f_{\mathrm 0}$ is still defined by the eq.~\ref{eq:HEW}. For small scattering angles ($\alpha \ll \theta_{\mathrm i}$), since we assumed a slow variation of $R_{\lambda}$ over angular scales of H/2 (and the same occurs for $K_{\lambda}$), we can approximate
\begin{equation}
	K_{\lambda}(\theta_{\mathrm i}, \theta_{\mathrm i}\pm \alpha) \approx R_{\lambda}(\theta_{\mathrm i})\pm \alpha \left. \frac{\partial K_{\lambda}(\theta_{\mathrm i}, \theta_{\mathrm s})}{\partial \theta_{\mathrm s}}\right|_{\theta_s = \theta_i}.
	\label{eq:pass_9}
\end{equation}
Substituting in the eq.~\ref{eq:pass_8}, we obtain
\begin{equation} 
\int_{f_0}^{\frac{2}{\lambda}}P(f)\,\mbox{d}f \approx \frac{\lambda^2\ln 2}{16\pi^2\sin^2\theta_{\mathrm i}}
	\label{eq:pass_10}
\end{equation}
because the two derivatives have opposite sign and cancel out. This is the same equation found for the case of a single-layer coating (eq.~\ref{eq:fstop}).

\section{Derivation of the PSD from the HEW for a fractal surface (single reflection)}
\label{sect:powerlawcalc}
We recall here the $H(\lambda)$ trend for a power-law PSD (eq.~\ref{eq:HEW_powerlaw}):
\begin{equation}
	H(\lambda) =2\left[\frac{16\pi^2 K_n}{(n-1)\ln 2}\right]^{\frac{1}{n-1}}\left(\frac{\sin\theta_{\mathrm i}}{\lambda}\right)^{\frac{3-n}{n-1}}
\label{eq:HEW_powerlaw_rec}
\end{equation}
we verify that it returns a power-law PSD if substituted in the differential eq.~\ref{eq:final}:
\begin{equation}
	\frac{P(f_{\mathrm 0})}{\lambda}\,\frac{\mbox{d}}{\mbox{d}\lambda}\!\left(\frac{H(\lambda)}{\lambda}\right) +\frac{\ln 2}{4\pi^2\sin^3\theta_{\mathrm i}} =0.
\label{eq:final_rec}
\end{equation}
To simplify the notation, we write simply $H$ instead of $H(\lambda)$: by carrying out the derivation,
\begin{equation}
	\frac{1}{\lambda}\,\frac{\mbox{d}}{\mbox{d}\lambda}\!\left(\frac{H}{\lambda}\right)=-\frac{4}{n-1}\left[\frac{16\pi^2 K_n}{(n-1)\ln2}\right]^{\frac{1}{n-1}}(\sin\theta_i)^{\frac{3-n}{n-1}}\lambda^{\frac{n-3}{n-1}-3}.
\label{eq:passage1}
\end{equation}
Using again the eq.~\ref{eq:HEW_powerlaw_rec}:
\begin{equation}
	\frac{1}{\lambda}\,\frac{\mbox{d}}{\mbox{d}\lambda}\!\left(\frac{H}{\lambda}\right)=-\frac{2H}{n-1}\lambda^{-3}
\label{eq:passage2}
\end{equation}
hence, the related PSD is
\begin{equation}
	P(f_0) = - \frac{\ln 2}{4\pi^2\sin^3\theta_{\mathrm i}} \left[\frac{1}{\lambda}\,\frac{\mbox{d}}{\mbox{d}\lambda}\!\left(\frac{H}{\lambda}\right)\right]^{-1}= \frac{\lambda^3\ln 2}{4\pi^2H\sin^3\theta_{\mathrm i}}\frac{n-1}{2}.
\label{eq:passage3}
\end{equation}
Now, we can derive $(n-1)/2$ from the eq.~\ref{eq:HEW_powerlaw_rec},
\begin{equation}
	\frac{n-1}{2}=\frac{4\pi^2H K_n}{\ln 2}\left(\frac{H}{2}\right)^{-n}\left(\frac{\sin\theta_i}{\lambda}\right)^{3-n}
\label{eq:passage4}
\end{equation}
and combining the eqs.~\ref{eq:passage3}-\ref{eq:passage4}, one obtains
\begin{equation}
	P(f_0) = K_n\left(\frac{H\sin\theta_i}{2\lambda}\right)^{-n},
\label{eq:passage5}
\end{equation}
that is, by recalling the eq.~\ref{eq:fstopfromH},
\begin{equation}
	P(f_0) = \frac{K_n}{f^n_{\mathrm 0}}
\label{eq:passage6}
\end{equation}
i.e., the expected power-law PSD.

\begin{acknowledgements}
Many thanks to G. Pareschi, O. Citterio, R. Canestrari, S. Basso, F. Mazzoleni, P. Conconi, V. Cotroneo (INAF/OAB) for support and useful discussions. The author is indebted to MIUR (the Italian Ministry for Universities) for the COFIN grant awarded to the development of multilayer coatings for X-ray telescopes. 
\end{acknowledgements}

\end{document}